\documentclass[aps,twocolumn,prd,showpacs,nofootinbib]{revtex4}

\usepackage{amsmath}
\usepackage{graphicx}
\usepackage{amssymb}

\newcommand{\sss}[1]{{\scriptscriptstyle{#1}}}

\newcommand{\nS}{n_{_{\mathrm{S}}}}

\newcommand{\nT}{n_{_{\mathrm{T}}}}

\newcommand{\uini}{\mathrm{ini}}
\newcommand{\uDBI}{\mathrm{\sss{DBI}}}
\newcommand{\usssS}{\mathrm{\sss{S}}}
\newcommand{\usz}{\mathrm{sz}}

\newcommand{\ub}{\mathrm{b}}
\newcommand{\us}{\mathrm{s}}

\newcommand{\udm}{\mathrm{dm}}

\newcommand{\cs}{c_\us}
\newcommand{\fNL}{f_{_\mathrm{NL}}}

\newcommand{\calP}{\mathcal{P}}

\newcommand{\CAMB}{\texttt{CAMB} }
\newcommand{\COSMOMC}{\texttt{COSMOMC} }

\newcommand{\OmegaCDM}{\Omega_\udm}
\newcommand{\OmegaB}{\Omega_\ub}

\newcommand{\mpl}{m_{_\mathrm{Pl}}}

\newcommand{\ie}{\textsl{i.e.}}

\newcommand{\order}[1]{\mathcal{O}\!\left(#1\right)}
\newcommand{\Hinf}{H_{\mathrm{inf}}}

\newcommand{\deriv}[2]{#1_{\negthinspace,#2}}
\newcommand{\dderiv}[3]{#1_{\negthinspace,#2#3}}

\newcommand{\ud}{\mathrm{d}}

\newcommand{\qmode}{\mu_\usssS}
\newcommand{\piv}{*}

\newcommand{\Pdbi}{P^\uDBI}

\newcommand{\Asz}{A_\usz}

\newcommand{\qmodeS}{\mu_\usssS}
\newcommand{\etapiv}{\eta_\piv}
\newcommand{\kpiv}{k_\piv}

\begin{document}

\title{Constraints on Kinetically Modified Inflation from WMAP5}

\author{Larissa Lorenz} \email{lorenz@iap.fr}
\author{J\'er\^ome Martin} \email{jmartin@iap.fr} \affiliation{
Institut d'Astrophysique de Paris, UMR 7095-CNRS, Universit\'e Pierre
et Marie Curie, 98bis boulevard Arago, 75014 Paris, France}

\author{Christophe Ringeval} \email{christophe.ringeval@uclouvain.be}
\affiliation{Theoretical and Mathematical Physics Group, Centre for
  Particle Physics and Phenomenology, Louvain University, 2 Chemin du
  Cyclotron, 1348 Louvain-la-Neuve, Belgium}

%\date{\today}
\date{July 15, 2008}

\begin{abstract}
  Single field inflationary models with a non-minimal kinetic term (also
  called k-inflationary models) can be characterised by the so-called
  sound flow functions, which complete the usual Hubble flow
  hierarchy. These parameters appear in the primordial power spectra of
  cosmological perturbations at leading order and, therefore, affect the
  resulting Cosmic Microwave Background (CMB) anisotropies. Using the
  fifth year Wilkinson Microwave Anisotropy Probe (WMAP5) data, we
  derive the marginalised posterior probability distributions for both
  the sound and Hubble flow parameters. In contrast to the standard
  situation, these parameters remain separately unbounded, and notably
  there is no longer any upper limit on $\epsilon_1$, the first Hubble
  flow function. Only special combinations of these parameters,
  corresponding to the spectral index and tensor-to-scalar ratio
  observables, are actually constrained by the data. The energy scale of
  k-inflation is nevertheless limited from above to $\Hinf \le 6 \times
  10^{-6} \mpl$ at two-sigma level. Moreover, for the sub-class of
  Dirac--Born--Infeld models, by considering the non-gaussianity bounds
  on the sound speed, we find a weak limit $\epsilon_1 < 0.08$ at $95\%$
  confidence level.

\end{abstract}

\pacs{98.80.Cq, 98.70.Vc}
\maketitle
%\tableofcontents

\section{Introduction}
\label{sec:intro}

Accelerated expansion of the early Universe in scalar field models is
usually achieved by domination of the field's potential over its kinetic
energy. However, it has been shown that inflation can also be supported
through modifications of the kinetic term, these models being known as
``k-inflation'' scenarios~\cite{ArmendarizPicon:1999rj}. Assuming the
gravity sector to be described by General Relativity [the signature of
the metric tensor being $(-,+,+,+)$], the action describing such a
''k-inflaton'' $\varphi(x^\mu)$ reads
\begin{equation}
  S=\dfrac{1}{2\kappa } \int \ud^4x \sqrt{-g} \left[ R + 2\kappa
    P\left(X,\varphi\right) \right],
\end{equation}
where $\kappa \equiv 8\pi/\mpl^2$ and where $X$ is defined by the
following expression
\begin{equation}
X \equiv -\dfrac{1}{2} g^{\mu \nu} 
\partial_\mu \varphi \partial_\nu \varphi .
\end{equation}
The quantity $P(X,\varphi)$ is any acceptable functional (see below) for
the scalar field Lagrangian. For instance, the canonical scalar field
model is recovered for $P(X,\varphi) = X - V(\varphi)$, where
$V(\varphi)$ is the scalar field potential. 

\par

The theory of k-inflation is interesting not only because it enlarges
the class of inflationary models that one can consider but also
because this is precisely this type of scenarios that emerges from
string theory. This is the case of the so-called Dirac--Born--Infeld
(DBI) scenarios of brane inflation, which have recently received
particular attention as effective field theory realizations of
inflation within string theory~\cite{Dvali:1998pa, Kachru:2003sx,
HenryTye:2006uv, Hertzberg:2007ke, Bean:2007eh, Peiris:2007gz,
Bean:2007hc, Lorenz:2007ze}. The DBI inflaton typically represents the
distance (\ie, an open string mode) between branes in a
higher-dimensional bulk, and the Lagrangian functional is given by the
determinant of the induced metric along the moving brane,
\begin{equation}
\label{eq:Pdbi}
\Pdbi(X,\varphi) = - T(\varphi) \sqrt{1 -\dfrac{2 X}{T(\varphi)}} 
+ T(\varphi)  - V(\varphi).
\end{equation}
Here, $T(\varphi)$ is a warping function encoding the bulk geometry, and
$V(\varphi)$ is the interaction potential between the branes, assembled
from contributions of various origin.

\par

In general, a model of k-inflation is sensible if the function
$P(X,\varphi)$ satisfies the two following conditions. Firstly,
$\deriv{P}{X}>0$, coming from the requirement that the Hamiltonian is
bounded from below and, secondly, $\deriv{P}{X}+2X
\dderiv{P}{X}{X}>0$, which is necessary in order for the Cauchy
problem to be well-posed, see for instance
Ref.~\cite{Bruneton:2007si}. Let us notice that the DBI models satisfy
these requirements. Moreover, these two conditions play an important
role at the perturbative level. Indeed, all kinetically modified
inflationary models share the common feature that the ``speed of
sound''
\begin{equation}
\label{eq:csdef}
\cs^2 \equiv \dfrac{\deriv{P}{X}}{\deriv{P}{X} + 2 X \dderiv{P}{X}{X}}\,,
\end{equation}is generically different from the speed of light and may be
space-time dependent. Clearly, the two above-mentioned
inequalities guarantee that $\cs^2>0$. This is relevant for the
treatment of cosmological perturbations. Indeed, for a flat
Friedmann-Lema\^{\i}tre--Robertson--Walker (FLRW) universe, it was
shown in Ref.~\cite{Garriga:1999vw}, that the Mukhanov--Sasaki mode
function $\qmode (k,\eta)= \zeta \sqrt{2\kappa \epsilon_1}/\cs$
($\zeta$ being the comoving curvature perturbation) satisfies the
modified equation of motion
\begin{equation}
\label{eq:qmode}
  \qmodeS'' + \left[\cs^2 k^2 - \dfrac{(a \sqrt{\epsilon_1}/\cs)''}{a 
      \sqrt{\epsilon_1}/\cs} \right] \qmodeS = 0.
\end{equation}
Derivatives are with respect to conformal time $\eta$, $a(\eta)$ is the
FLRW scale factor and $\epsilon_1 = -\ud \ln H/\ud \ln a$ is the first
Hubble flow function ($H$ being the Hubble parameter). In addition to
the induced modification of the perturbations' effective potential,
$\cs$ also appears as their propagation speed. Let us notice that for
DBI inflation, Eqs.~(\ref{eq:Pdbi}) and~(\ref{eq:csdef}) give
\begin{equation}
\label{eq:csdbi}
\cs = \sqrt{1 - \dfrac{{\varphi'}^2}{a^2 T(\varphi)}}\, ,
\end{equation}
and therefore ensure that $\cs\le1$. In the following, we will only
consider the k-inflation models for which $\cs \le 1$ (see, however,
Refs.~\cite{Bruneton:2006gf, Bonvin:2006vc,
Ellis:2007ic,Babichev:2007dw, Bonvin:2007mw, Kang:2007vs}).

\par

In this paper, we are interested in constraining the shape of the
scalar and tensor primordial power spectra for all the kinetically
modified models admitting an observable ``slowly-evolving'' phase. To
this end, we assume the existence of a background solution supporting
a long enough inflationary era to solve the FLRW problems.

\par

The fact that $\cs$ is time-dependent and appears in front of the
$k^2$ term in Eq.~(\ref{eq:qmode}) has important consequences and
brings new complications into the problem. In particular, one can no
longer use the standard techniques consisting in solving the mode
equation in terms of Bessel functions. In the case of k-inflation, one
is therefore forced to rely on a new method of approximation. This is
done in Ref.~\cite{Kinney:2007ag} and in Ref.~\cite{Lorenz:2008xx},
this last article making use of the Wentzel--Kramers--Brillouin
(WKB)/uniform approximation~\cite{Habib:2002yi,
  Habib:2004kc,Martin:2002vn}. In this manner, one can solve
Eq.~(\ref{eq:qmode}) perturbatively in terms of a double hierarchy of
functions~\cite{Alishahiha:2004eh, Chen:2006nt, Kinney:2007ag,
  Peiris:2007gz, Lorenz:2008xx} defined by
\begin{eqnarray}
\label{eq:hierarchies}
\begin{aligned}
  \epsilon_{i+1} & = \dfrac{\ud \ln |\epsilon_i|}{\ud \ln a}\,, \quad
  \epsilon_0 = \dfrac{H_\uini}{H}\,, \\
  \delta_{i+1} & = \dfrac{\ud \ln |\delta_i|}{\ud \ln a}\,, \quad
  \delta_0 = \dfrac{{\cs}_\uini}{\cs}\,.
\end{aligned}
\end{eqnarray}
The $\epsilon_i$ are the usual Hubble flow functions introduced in
Refs.~\cite{Schwarz:2001vv,Leach:2002ar,Schwarz:2004tz}, while the
$\delta_i$ are their equivalent based on the sound speed and therefore
encode its rate of change. At first order in these parameters, the
scalar and tensor primordial power spectra\footnote{For single field
models, $\zeta$ is a conserved quantity on ``super-sonic'' length
scales. As in the standard case~\cite{Martin:1997zd}, this allows us to
propagate the primordial spectrum from the end of inflation to the time
of decoupling.} can be expressed as~\cite{Lorenz:2008xx}
\begin{equation}
\label{eq:powerspectra}
\begin{aligned}
  \calP_\zeta & = \dfrac{\Hinf^2}{\pi \mpl^2 \epsilon_1 \cs} \left[1 -
    2(D+1) \epsilon_1 - D \epsilon_2 + (D+2) \delta_1
    \phantom{\dfrac{k}{\kpiv}}\right. \\ & - \left. \left(2 \epsilon_1
      + \epsilon_2 - \delta_1 \right) \ln \dfrac{k}{\kpiv} \right], \\
  \calP_h & = \dfrac{16 \Hinf^2}{\pi \mpl^2} \left[1 - 2(D+1-\ln \cs)
    \epsilon_1 -2 \epsilon_1 \ln \dfrac{k}{\kpiv} \right].
\end{aligned}
\end{equation}
The wavenumber $\kpiv$ is a pivot scale around which these power spectra
are expanded, see Ref.~\cite{Lorenz:2008xx}. All the functions appearing
in this equation have to be evaluated at the conformal time $\etapiv$ at
which this pivot mode crossed the sound horizon, namely the solution of
\begin{equation}
  \kpiv \etapiv  = - \dfrac{1}{\cs(\etapiv)}\,.
\end{equation}
Notice that although the tensor perturbations are not affected by the
non-minimal kinetic terms in the scalar sector, $\cs$ nevertheless
appears in $\calP_h$ due to the evaluation of the flow functions
at the sound horizon crossing time instead of Hubble crossing. In
Eq.~(\ref{eq:powerspectra}), $D = 1/3 - \ln 3\simeq -0.76$ is a
constant coming from the uniform approximation, and an overall
constant factor of $18 \exp(-3)$ has been absorbed into
$\Hinf^2$~\cite{Martin:2002vn}. From Eq.~(\ref{eq:powerspectra}) we
can read off the spectral indices to first order,
\begin{equation}
\label{eq:indices}
\begin{aligned}
\nS-1&=-2\epsilon_{1}-\epsilon_{2}+\delta_{1}\, ,\quad
\nT=-2\epsilon_{1},
\end{aligned}
\end{equation}
and the k-inflation tensor to scalar ratio $r=16 \epsilon_1 \cs$.

\par

As expected, the power spectra in Eq.~(\ref{eq:powerspectra}) are not
the same than those obtained for a canonically normalised scalar field
due to the presence of the two sound flow parameters $\cs$ and
$\delta_1$. Notice that in the limit $\cs = 1 - \order{\epsilon_i}$,
one has $\delta_1 = \order{\epsilon_i^2}$ and the above spectra only
differ from their usual analogue at second order in the Hubble flow
functions~\cite{Steer:2003yu}.

\par

In the following, we compare the induced Cosmic Microwave Background
(CMB) anisotropies seeded by these power spectra to the WMAP5
data~\cite{Gold:2008kp, Hill:2008hx, Hinshaw:2008kr, Nolta:2008ih,
Dunkley:2008ie, Komatsu:2008hk}. Using Markov--Chains--Monte--Carlo
(MCMC) methods, we extract the posterior marginalised probability
distributions on the sound and Hubble flow parameters and discuss the
physical consequences. Moreover, as shown in Refs.~\cite{Seery:2005wm,
Chen:2006nt, Bean:2008ga, Gauthier:2008mq}, kinetically modified
inflation models may induce a large amount of non-gaussianities in the
statistics of the CMB anisotropies, usually measured in terms of the
$\fNL$ parameter~\cite{Komatsu:2001rj}. The expected value for $\fNL$
strongly depends on the model at hand: as shown in
Ref.~\cite{Chen:2006nt}, $\fNL$ for equilateral configurations demands
the knowledge of the third $X$-derivative of $P(X,\varphi)$ in addition
to the sound and Hubble flow functions. For these reasons, we have
chosen to split our analysis into two parts. In the first section, we do
not consider the non-gaussianity constraints, and our results are only
based on the observed two-point functions of the CMB temperature and
polarisation anisotropies. In the second part, we implement the WMAP5
bounds on $\fNL$ as prior knowledge on the sound speed values in the
restricted class of DBI models: for those, we indeed have the universal
expression~\cite{Chen:2006nt,Gao:2008dt, Langlois:2008qf}
\begin{equation}
\label{eq:fnl}
  \fNL=\dfrac{35}{108}\left(1-\dfrac{1}{\cs^2}\right ) 
  + \order{\epsilon_i,\delta_i}.
\end{equation}
{}From now on, $\fNL$ is implicitly assumed to refer to equilateral
configurations, the ``local'' estimator being $\fNL^{\mathrm{local}}
\sim \order{\epsilon_i,\delta_i}$ and well below the current detection
thresholds, see Ref.~\cite{Chen:2006nt, Komatsu:2008hk, Bean:2008ga}.

\section{CMB power spectra}
\label{sec:cmbspectra}

We have used a modified version of the \CAMB code~\cite{Lewis:1999bs} to
compute the CMB temperature and polarisation anisotropies seeded by the
scalar and tensor primordial power spectra of
Eqs.~(\ref{eq:powerspectra}). Comparison with the WMAP data is then
performed by exploring the full parameter space, including the standard
cosmological parameters, through MCMC Bayesian methods implemented in
the \COSMOMC package~\cite{Lewis:2002ah, Trotta:2008qt}. The likelihood
estimator is provided by the WMAP team~\cite{Dunkley:2008ie}.

\subsection{Cosmological model}

\begin{figure}
\begin{center}
\includegraphics[width=0.5\textwidth]{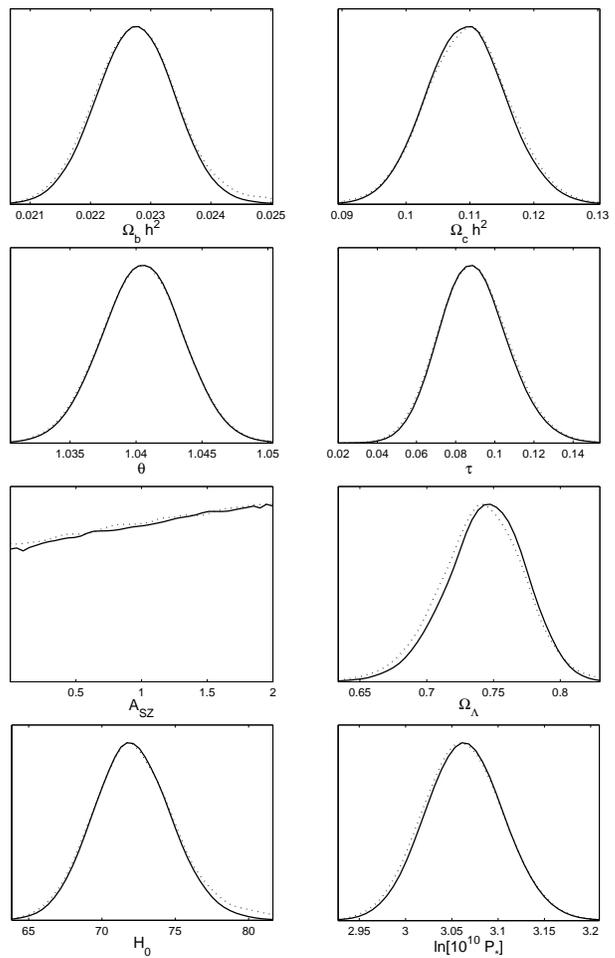} \caption{One
dimensional marginalised posterior probability distribution (solid) and
mean likelihoods (dotted) for the base and derived cosmological
parameters, given the WMAP5 data and the k-inflation power
spectra of Eq.~(\ref{eq:powerspectra}).}  \label{fig:dbicosmo}
\end{center}
\end{figure}

The assumed cosmological model is a flat $\Lambda$CDM universe involving
four cosmological base parameters (with priors chosen from the
posteriors of the previous CMB experiments): the number density of
baryons $\OmegaB$, of cold dark matter $\OmegaCDM$, the reionization
optical depth $\tau$, and $\theta$ which measures the ratio of the
acoustic horizon to the angular diameter distance (see
Ref.~\cite{Lewis:2002ah}). As performed by the WMAP team, we have now
included the lensing of the CMB power spectra and performed a
marginalisation over the Sunyaev-Zel'dovich amplitude $\Asz$, in unit of
the analytical model of Ref.~\cite{Komatsu:2002wc} and with a flat prior
in $[0,2]$. For the sake of clarity, and to allow comparison with
previous works~\cite{Martin:2006rs}, we have, at this stage, not
considered more data sets than the WMAP5 data, the Hubble Space
Telescope (HST) constraint ($H_0 = 72 \pm 8\,
\mathrm{km/s/Mpc}$~\cite{Freedman:2000cf}) and a top hat prior on the
age of the universe between $10\,\mathrm{Gyrs}$ and $20
\,\mathrm{Gyrs}$. Notice that we have chosen the pivot scale to be in
the middle of the observable range, namely
$\kpiv=0.05\,\mathrm{Mpc}^{-1}$~\cite{Liddle:2006ev}.

\subsection{Observable primordial parameters and priors}

It is important to notice that the new parameters appearing in the
primordial power spectra of Eqs.~(\ref{eq:powerspectra}) induce new
degeneracies. In particular, and contrary to the usual case,
$\epsilon_1$ is no longer uniquely linked to the tensor to scalar ratio
due to the presence of $\cs$. Therefore, the product $\epsilon_1 \cs$ is
a better motivated observable parameter in k-inflation. Its order of
magnitude being a priori unknown, we have chosen a uniform logarithmic
prior\footnote{Recall that we are considering the case $\cs^2 \le 1$.}:
$-10 <\log(\epsilon_1 \cs) < -0.5$ (Jeffreys' prior).

\par

Apart from the overall amplitude, $\calP_\zeta(k)$ in
Eqs.~(\ref{eq:powerspectra}) has exactly the same functional form than
the canonical scalar slow-roll power spectrum at first order upon the
replacement $\epsilon_1 \rightarrow \epsilon_1 - \delta_1$, and
$\epsilon_2 \rightarrow \epsilon_2 + \delta_1$. In particular, this is
the case for the spectral index $\nS$ in Eq.~(\ref{eq:indices}). As a
result, $\epsilon_1$, $\epsilon_2$ and $\delta_1$ are fully degenerated
and there are only two observable parameters, namely $(\epsilon_1
-\delta_1)$ and $(\epsilon_2 + \delta_1)$, for which we have chosen a
uniform prior in $[-0.3, 0.3]$. This important point already shows that,
at the scalar perturbation level, there is no chance of constraining
$\delta_1$ separately, neither $\epsilon_1$ nor $\epsilon_2$; this is
significantly different from the usual slow-roll
constraints~\cite{Barger:2003ym, Leach:2003us, Martin:2006rs,
Finelli:2006fi}. However, considering the primordial tensor power
spectrum $\calP_{h}(k)$, one sees that $\epsilon_1$ is still uniquely
linked with the tensor spectral index $\nT$ [see
Eq.~(\ref{eq:indices})]. As one may expect, it is unconstrained by the
data at present, and so should be $\epsilon_1$. We have however included
it for completeness as the third primordial parameter in the MCMC
exploration with a logarithmic prior in $-5 \le \log\epsilon_1 \le
-0.5$. Finally, the overall amplitude of the CMB anisotropies is fixed
by $P_\piv = \calP_\zeta(k_\piv)$ for which we have chosen the usual
uniform prior distribution on $\ln(10^{10}P_\piv)$ in $[2.7,4]$.

\subsection{Results}

\begin{figure}
\begin{center}
  \includegraphics[width=0.5\textwidth]{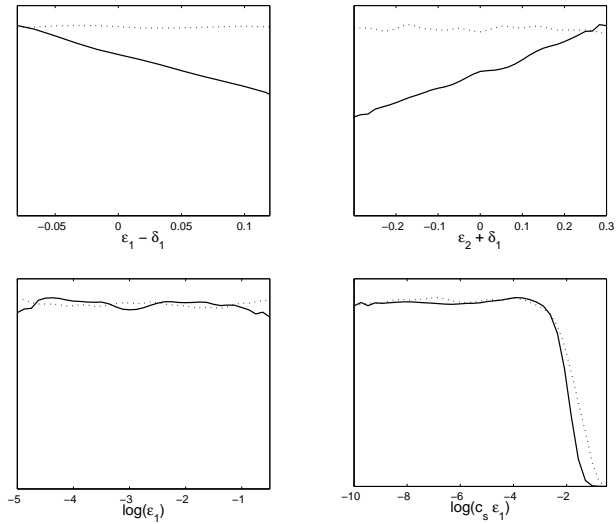} \caption{One
    dimensional marginalised posterior probability distribution (solid)
    and mean likelihood (dotted) for the sampled sound and Hubble flow
    parameters given the WMAP5 data. Only $\log(\epsilon_1\cs )$ is
    upper limited, this parameter fixing the tensor-to-scalar
    ratio. Notice in particular that $\epsilon_1$ is not
    constrained. The parameters $(\epsilon_1-\delta_1)$ and $(\epsilon_2
    + \delta_1)$ are, however, correlated (see
    Fig.~\ref{fig:dbiprim2D}).}  \label{fig:dbiprim}
\end{center}
\end{figure}

\begin{figure}
\begin{center}
  \includegraphics[width=0.5\textwidth]{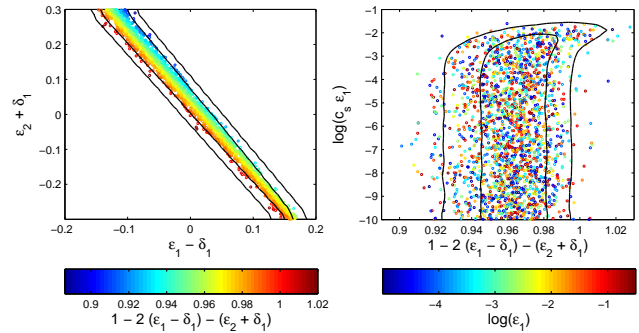} \caption{Two
    dimensional marginalised posteriors (point density) together with
    the one and two-sigma contours. Although $(\epsilon_1 - \delta_1)$
    and $(\epsilon_2 + \delta_1)$ are not bounded, they are strongly
    correlated to give acceptable values of the spectral index (left
    panel). The right panel exhibits the correlations between
    $\log(\epsilon_1\cs )$ and $\nS=1-2(\epsilon_1-\delta_1)
    -(\epsilon_2 + \delta_1)$.}  \label{fig:dbiprim2D}
\end{center}
\end{figure}

The MCMC exploration has been stopped according to the R--statistics
implemented in \COSMOMC~\cite{Gelman:1992, Lewis:2002ah} as soon as the
variances between the different chains agree at a few percent. The
chains contain $800\,000$ samples and lead to the marginalised posterior
probability distributions represented in Fig.~\ref{fig:dbicosmo} to
Fig.~\ref{fig:dbiprim2D}. The base cosmological parameters associated
with the $\Lambda$CDM universe are compatible with the recent
results~\cite{Komatsu:2008hk}, although our primordial power spectra are
different.

Fig.~\ref{fig:dbiprim} shows the posterior probability distributions
(solid curves) for each of the primordial parameters we are interested
in. As we expected, there is no bound on $\log \epsilon_1$ and the
posterior is just given by the prior. On the other hand, the maximum
amount of tensor modes in the CMB power spectra gives an upper limit on
the parameter $\log(\epsilon_1\cs)$. Its posterior is indeed upper
limited and at $95 \%$ of confidence, one finds
\begin{equation}
\label{eq:cseps1twosig}
\log \left(\epsilon_1\cs \right)  \le -2.3\, .
\end{equation}
Notice that the above upper limit is stronger than the ones reported
in the recent literature~\cite{Komatsu:2008hk, Kinney:2008wy}. This
effect comes from our Jeffreys' prior choice on $\epsilon_1\cs $,
which favours low values of the parameter~\cite{Trotta:2008qt}, and
possibly from the new natural choice of the sampling parameters
$(\epsilon_1 - \delta_1)$ and $(\epsilon_2 + \delta_1)$. Let us
mention that this prior choice is motivated by our ignorance on the
scale at which tensor perturbations contribute to the CMB, a
priori\footnote{From a flat prior on $\epsilon_1\cs$, we obtain
  $\epsilon_1 \cs < 0.023 $ at two-sigma level (by importance
  sampling).}. The slight discrepancies appearing in between the
posterior and the mean likelihood for this parameter signal some
correlations with other parameters~\cite{Lewis:2002ah,
  Martin:2004yi}. Also, both $(\epsilon_1-\delta_1)$ and $(\epsilon_2
+ \delta_1)$ exhibit unbounded distributions, and their mean
likelihood (dotted curves) differs from the marginalised
posterior. The two-dimensional marginalised probability distribution
in the plane $(\epsilon_1-\delta_1,\epsilon_2 + \delta_1)$ is plotted
in Fig.~\ref{fig:dbiprim2D} and clearly exhibits the above-mentioned
correlations. The physical interpretation is straightforward: the
``correlation strip'' is given by the allowed values of the scalar
spectral index $\nS$ [see Eq.~(\ref{eq:indices})]. The barely visible
thickening of this strip comes from some slight correlations with the
parameter $P_\piv$, and then $\Asz$, due to the functional form of the
power spectra. Finally, as can be seen in Fig.~\ref{fig:dbiprim2D},
large values of $\log( \epsilon_1\cs)$ are also degenerated with the
spectral index $\nS$ in a way reminiscent of the corresponding
correlation between tensor-to-scalar ratio and spectral index in
standard inflation.

\begin{figure}
\begin{center}
\includegraphics[width=0.5\textwidth]{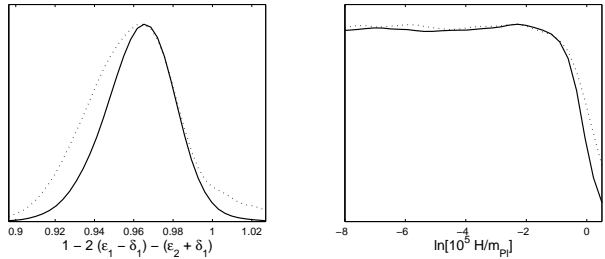}
\caption{Marginalised posterior probability distribution for the derived
primordial parameters $\nS$ and $\Hinf$ obtained by importance
sampling.}  \label{fig:dbiprimderiv}
\end{center}
\end{figure}

In Fig.~\ref{fig:dbiprimderiv}, we have used importance sampling to
reconstruct the probability distribution of $\nS$ and $\ln(10^5
\Hinf/\mpl)$. The probability distribution for $\nS$ ends up being
compatible with the one found by assuming a power law primordial
spectrum~\cite{Komatsu:2008hk}. At $95\%$ of confidence, one gets
\begin{equation}
\label{eq:nstwosig}
0.003 \le 2 (\epsilon_1-\delta_1) + (\epsilon_2 + \delta_1) \le 0.075.
\end{equation}
This limit favours a ``red tilted'' scalar power spectrum, at two-sigma
level, in presence of tensor modes and varying sound speed\footnote{This
is, however, prior dependent; from a flat prior on $\epsilon_1\cs $, the
bounds become $[-0.024,0.063]$.}. The probability distribution
associated with $\Hinf$ is only upper limited since this parameter is
fully degenerated with $\epsilon_1\cs $ in fixing the amplitude of the
primordial scalar power spectrum [see Eq.~(\ref{eq:powerspectra})]. The
prior distribution on $\Hinf$ is logarithmically uniform and induced by
our Jeffreys' prior on $ \epsilon_1\cs$ and $P_\piv$. This is convenient
since the energy scale of inflation is also unknown a priori. We find
\begin{equation}
\label{eq:hinftwosig}
\ln \left(10^5 \dfrac{\Hinf}{\mpl} \right) \le -0.59 ,
\end{equation}
at $95\%$ of confidence.

\section{Including non-gaussianity}

In this section, we use the two-sigma WMAP5 bound on the $\fNL$
parameter associated with ``equilateral configurations'' in Fourier
space~\cite{Komatsu:2008hk},
\begin{equation}
\label{eq:fnltwosig}
-151 \le \fNL \le 253.
\end{equation}
As already mentioned, the predicted value of $\fNL$ for k-inflation
models is not universal and depends on the explicit form of the third
derivative of $P(X,\varphi)$ with respect to $X$. As a result, we have
chosen to limit our analysis to the sub-class of DBI models: for those,
Eq.~(\ref{eq:fnl}) can be used.

\subsection{New priors}

A straightforward implementation of the non-gaussianity bound is to
include it as prior knowledge on the acceptable $\cs$ values. In fact,
this occurs via the so-called ``Lorentz factor'' $\gamma^2 \equiv
1/\cs^2$ which enters into the expression (\ref{eq:fnl}). Since its
order of magnitude is now fixed, we have chosen to sample the parameter
space over $\gamma^2$, instead of $\log(\epsilon_1\cs )$, with an
uniform prior over $[1,467]$. All the other sampling parameters and
priors remain the same as in Sect.~\ref{sec:cmbspectra}, as well as the
cosmological model.

\subsection{Results}

\begin{figure}
\begin{center}
\includegraphics[width=0.5\textwidth]{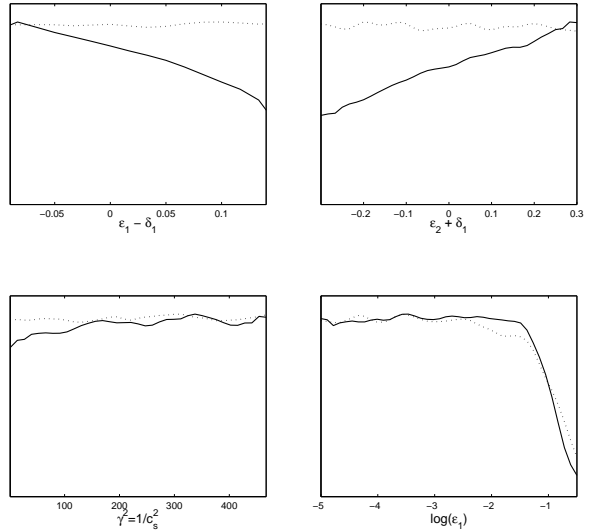}
\caption{Marginalised posterior probability distribution for the DBI
primordial parameters by including the non-gaussianity bounds on
$\fNL$. The first Hubble flow parameter $\epsilon_{1}$ is now upper
limited.}  \label{fig:dbingprim}
\end{center}
\end{figure}

The marginalised posterior probability distributions associated with the
base and derived cosmological parameters are essentially the same as in
the previous section and we do not discuss them any further (see
Fig.~\ref{fig:dbicosmo}). The posteriors of the parameters appearing in
the primordial power spectra are plotted in Fig.~\ref{fig:dbingprim}.

\begin{figure}
\begin{center}
\includegraphics[width=0.5\textwidth]{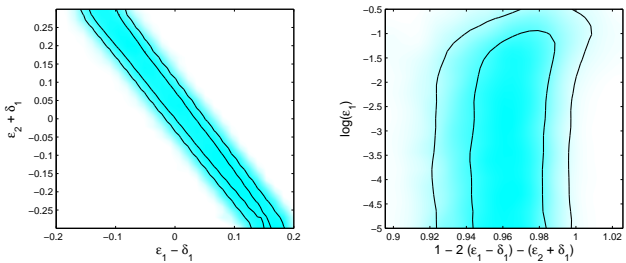} \caption{One and
two-sigma contours of the two dimensional marginalised posteriors for
correlated pairs of primordial parameters. The shading traces the
two-dimensional mean likelihood. The right panel shows that the
non-gaussianity bounds on $\fNL$ induces an upper bound on the first
Hubble flow parameter $\epsilon_{1}$.}  \label{fig:dbingprim2D}
\end{center}
\end{figure}

In fact, the non-gaussianity prior imposed on $\gamma^2$ breaks the
degeneracies we encountered previously between $\epsilon_1$ and $\cs$,
and which motivated the use of $\log(\epsilon_1\cs )$ as a sampling
parameter. The sound speed being now limited from below, the
tensor-to-scalar ratio becomes sensitive to the first Hubble flow
parameter and we find a ``weak'' upper bound on $\log \epsilon_1$. At
$95\%$ of confidence,
\begin{equation}
\label{eq:eps1twosig}
\log \epsilon_1 \le -1.1\, .
\end{equation}
On the other hand, $(\epsilon_1 - \delta_1)$ and $(\epsilon_2 +
\delta_1)$ are still unbounded but correlated along the acceptable
values of the spectral index (see Fig.~\ref{fig:dbingprim2D}). As a
result, $\delta_1$ remains unconstrained, and so does $\epsilon_2$. The
posteriors for the spectral index and the energy scale of inflation
$\Hinf$ appear to be relatively insensitive to the inclusion of the
non-gaussianity bound. Their respective two-sigma limits are still given
by Eq.~(\ref{eq:nstwosig}) and Eq.~(\ref{eq:hinftwosig}).

\par

In addition to imposing strong bounds on $\cs$ for DBI models, the
non-gaussianity limits of Eq.~(\ref{eq:fnltwosig}) allow
disambiguation of the first Hubble flow parameter, and thus, some
inference on the evolution of the Hubble parameter during
inflation. It is clear that the limit of Eq.~(\ref{eq:eps1twosig}) is
not very tight, and quite close to the values acceptable for the
``slowly-evolving'' limit. However, improving the bounds on $\fNL$
will directly improve the determination of the acceptable
$\log(\epsilon_1)$ values: one may therefore expect tighter
constraints from the next generation of CMB experiments.

\section{Conclusion}

We have used the WMAP fifth year data to constrain all single field
k-inflationary models admitting a phase of ``slow-rolling'' evolution
during the generation of the cosmological perturbations.

\par

The probability distributions of the Hubble and sound flow parameters,
$\epsilon_i$ and $\delta_i$, evaluated when the scalar perturbations
crossed the sound horizon, are significantly different than the ones
associated with canonically normalised single field models. The presence
of two additional parameters in the primordial power spectra, $\cs$ and
$\delta_1$, leads to new degeneracies in the parameter space: there is
no longer any constraint on $\epsilon_1$ and $\epsilon_2$ separately,
neither on $\cs$ or $\delta_1$ alone. The energy scale of inflation is
nevertheless upper limited: $\Hinf/\mpl \le 6\times 10^{-6}$ at $95\%$
of confidence. In fact, only two particular combinations of the
primordial parameters end up being constrained by the CMB power spectra:
the quantity $\epsilon_1\cs$ in Eq.~(\ref{eq:cseps1twosig}), which
encodes the tensor-to-scalar ratio in the CMB spectra, and
$2(\epsilon_1-\delta_1) + (\epsilon_2 + \delta_1)$, which fixes the
scalar spectral index in Eq.~(\ref{eq:nstwosig}). Under our prior
assumptions, Eq.~(\ref{eq:nstwosig}) shows that the data favour a
``red-tilted'' scalar power spectrum, at two-sigma level, with tensor
modes and varying sound speed.

\par

Implementing the current WMAP5 bound on non-gaussianity through the
$\fNL$ parameter (for equilateral configurations) breaks some of the
above-mentioned degeneracies. For the DBI inflationary models, more than
restricting the acceptable values of the sound speed $\cs$, we find that
the non-gaussianity bound induces an upper limit on the first Hubble
flow parameter: $\log(\epsilon_1) < -1.1$ at two-sigma confidence level.

\par

The present results could be improved by various means. A first step may
be to add more data sets, but according to the previous discussion, this
will certainly improve only the limits on the spectral index and
tensor-to-scalar ratio combinations without breaking the degeneracies
between the $\epsilon_i$ and $\delta_i$. Therefore, it is clearly more
promising to improve the tests that bring new information on
non-gaussianities. We have implemented them for DBI models only through
a prior knowledge over $\cs$. Clearly, our approach could be improved by
directly evaluating the three-point functions and comparing it to the
observed one, provided one can define an tractable likelihood for MCMC
exploration. Such an approach is currently technically
difficult. Evaluating the three-point functions from CMB data is not
trivial (see Appendix A in Ref.~\cite{Komatsu:2008hk}); and, from a
theoretical point of view, new parameters have to be introduced to
quantify the non-gaussian signals in a generic way for
k-inflation. Therefore, one faces the risk to add even more
degeneracies. On the other hand, with the incoming flow of data on
cosmological observables, this represents an interesting challenge for
the forthcoming years.

\begin{acknowledgements}
  We would like to thank G.~Geshnizjani for useful discussions. The
  computations have been performed on the PLANCK--HFI French
  Processing Center hosted at the Institut d'Astrophysique de
  Paris. This work is partially supported by the Belgian Federal
  Office for Science, Technical and Cultural Affairs, under the
  Inter-university Attraction Pole grant P6/11. LL acknowledges
  support through a DAAD PhD scholarship.

\end{acknowledgements}

\bibliography{references}
\end{document}